\documentclass[preprint,showkeys,showpacs,superscriptaddress,nofootinbib]{revtex4}
  \usepackage{graphicx,subfigure}
  \usepackage{amsmath}
   \usepackage{color}
  \begin{document}
  \title{Study of nonleptonic $B_{q}^{\ast}$ ${\to}$ $D_{q}V$ and $P_{q} D^*$ weak decays}
  \author{Qin Chang}
  \affiliation{Institute of Particle and Nuclear Physics,
              Henan Normal University, Xinxiang 453007, China}
  \affiliation{Institute of Particle Physics,
              Central Normal University, Wuhan 430079, China}
  \affiliation{State Key Laboratory of Theoretical Physics,
               Institute of Theoretical Physics,
               Chinese Academy of Sciences, Beijing 100190, China}
  \author{Xiaohui Hu}
  \affiliation{Institute of Particle and Nuclear Physics,
              Henan Normal University, Xinxiang 453007, China}
  \author{Junfeng Sun}
  \affiliation{Institute of Particle and Nuclear Physics,
              Henan Normal University, Xinxiang 453007, China}
  \author{Xiaolin Wang}
  \affiliation{Institute of Particle and Nuclear Physics,
              Henan Normal University, Xinxiang 453007, China}
  \author{Yueling Yang}
  \affiliation{Institute of Particle and Nuclear Physics,
              Henan Normal University, Xinxiang 453007, China}
  \begin{abstract}
  Motivated by the powerful capability of measurement for the
  $b$-flavored hadron rare decays at LHC and SuperKEKB/Belle-II, the
  nonleptonic $\bar{B}^{\ast}$ ${\to}$ $D\bar{D}^{\ast}$,
  $D{\rho^-}$, $DK^{\ast-}$, ${\pi}D^{\ast}$ and $KD^{\ast}$
  weak decays are studied in detail.
  With the amplitudes calculated with factorization
  approach and the form factors of $B^{\ast}$ transition
  into pseudoscalar meson evaluated with the BSW model,
  branching fractions and polarization fractions are
  firstly presented. Numerically,
  the CKM-favored $\bar{B}_{q}^{\ast}$ ${\to}$
  $D_{q}D_{s}^{{\ast}-}$ and $D_{q}{\rho}^{-}$
  decays have large branching fractions, $\sim$
  $10^{-8}$, which should be sought for with
  priority and firstly observed by LHC and Belle-II
  experiments.
  The $\bar{B}^{\ast}_q$ ${\to}$ $D_qK^{\ast}$ and $D_q{\rho}$
  decays are dominated by the longitudinal polarization
  states. While, the parallel polarization fractions
  of $\bar{B}^{\ast}_q$ ${\to}$ $D_q\bar{D}^{\ast}$ decays
  are comparable with the longitudinal ones,
  numerically, $f_{\parallel}$ $+$ $f_{L}$ ${\simeq}$ 95\%
  and $f_{L}:f_{\parallel}$ $\simeq$ $5:4$. Some comparisons between $\bar{B}^{*0}_q$ $\to$ $D_q V$ and their corresponding $\bar{B}^{0}_q$ $\to$ $D^*_q V$ decays are performed, and the relation $ f_{L,\parallel}(\bar{B}^{\ast 0}\to D V)\simeq f_{L,\parallel}(\bar{B}^0\to D^{\ast +} V^-) $ is presented.
  Besides, with the implication of $SU(3)$ flavor symmetry,
  some useful ratios $ R_{\rm du}$ and $ R_{\rm ds}$
   are discussed in detail, and suggested to be verified experimentally.
  \end{abstract}
  \maketitle
  \section{Introduction}
  \label{sec01}
  The b physics plays an important role in testing the flavor
  dynamics of Standard Model (SM), exploring the source of
  $CP$ violation, searching the indirect hints of new physics,
  investigating the underling mechanisms of QCD {\em et al.},
  and thus attracts much experimental and theoretical
  attention.
  With the successful performance of BABAR, Belle, CDF and D0
  in the past years, many $B_{u,d,s}$ meson decays have
  been well measured.
  Thanks to the ongoing LHCb experiment \cite{ref:LHCb2013}
  at LHC and forthcoming Belle-II experiment
  \cite{ref:BelleII2011} at SuperKEKB,
  experimental analysis of $B$ meson decays is entering
  a new frontier of precision.
  By then, besides $B_{u,d,s}$ mesons, the rare decays
  of some other $b$-flavored hadrons are hopefully to
  be observed, which may provide much more extensive
  space for $b$ physics.

  The excited states $B^{\ast}_{u,d,s}$ with
  quantum number of $n^{2s+1}L_{J}$ $=$ $1^{3}S_{1}$
  and $J^{P}$ $=$ $1^{-}$ ( $n$, $L$, $s$, $J$
  and $P$ are the quantum numbers of radial,
  orbital, spin, total angular momenta and
  parity, resptctively), which will be referred
  as $B^{\ast}$ in this paper, had been observed
  by CLEO, Belle, LHCb and so on \cite{PDG14}.
  However, except for their masses, there is
  no more experimental information due to
  the fact that the production of $B^{\ast}$
  mesons are mainly through $\Upsilon(5S)$
  decays at $e^{+}e^{-}$ colliders and
  the integrated luminosity is not high
  enough for probing the $B^{\ast}$ rare
  decays.
  Moreover, $B^{\ast}$ decays are dominated by
  the radiative processes $B^{\ast}$ $\to$
  $B{\gamma}$, and the other decay modes are
  too rare to be measured easily.
  Fortunately, with annual integrated luminosity
  ${\sim}$ $13$ ${\rm ab}^{-1}$ \cite{ref:BelleII2011}
  and the cross section of $\Upsilon(5S)$ production
  in $e^{+}e^{-}$ collisions
  ${\sigma}(e^{+}e^{-}{\to}\Upsilon(5S))$ $=$
  $(0.301{\pm}0.002{\pm}0.039)$ nb \cite{Huang:2006mf},
  it is expected that about $4{\times}10^{9}$
  $\Upsilon(5S)$ samples could be produced per year
  at the forthcoming super-B factory SuperKEKB/Belle-II, which implies that the $B^{\ast}$ rare decays with branching fractions $\gtrsim 10^{-9}$ are possible to be observed.
  Besides, due to the much larger production
  cross section of $pp$ collisions, experiments at
  LHC \cite{Aaij:2010gn,Aaij:2014jba}
  also possibly provide some experimental information
  for $B^{\ast}$ decays.

  With the rapid development of experiment, accordingly,
  the theoretical evaluations for $B^{\ast}$ weak decays
  are urgently needed and worthful.
  Nonleptonic $B^{\ast}$ weak decays allow one to
  overconstrain parameters obtained from $B$ meson
  decay, test various models and improve our understanding
  on the strong interactions and the mechanism responsible
  for heavy meson weak decay.
  The observation of an anomalous production rate
  of $B^{\ast}$ weak decays would be a hint of
  possible new physics beyond SM.
  In addition, the $B^{\ast}$ weak decay porvide
  one unique opportunity of observing the weak
  decay of a vector meson, where polarization
  effects can be used as tests of the underlying
  structure and dynamics of hadrons.
  To our knowledge, few previous theoretical works
  come close to studying $B^{\ast}$ weak decays.
 Compared with the $B^{*}\to PP$ decays, which are suppressed dynamically by the orbital angular momentum of final states, $B^{*}\to PV$ decays are expected to have much larger branching fractions, and hence generally much easier to be measured. So, in this paper, we will estimate the observables of
  nonleptonic two-body $B^{*}\to PV$ weak decay to offer a ready
  reference.

  Our paper is organized as follows.
  In section \ref{sec02}, after a brief review of
  the effective Hamiltonian and factorization approach,
  the explicit amplitudes of $B_{u,d,s}^{\ast}$
  ${\to}$ $D_{u,d,s}^{(\ast)}M$ decays are calculated.
  In sections \ref{sec03}, the numerical results and
  discussions are presented.
  Finally, we summarize in section \ref{sec04}.

  \section{Theoretical Framework}
  \label{sec02}
  Within SM, the effective Hamiltonian responsible for
  nonleptonic $B^{\ast}$ weak decay is
  \cite{ref:Buras1}
  \begin{equation}
 {\cal H}_{\rm eff} =
  \frac{G_F}{\sqrt{2}} \sum_{q, q^{\prime} = u, c}
  \Big[ V_{qb}V_{q^{\prime}p}^{\ast}
  \sum_{i = 1}^{2} C_{i}({\mu}) O_{i} ({\mu})
 +V_{qb}V_{qp}^{\ast}
  \sum_{i = 3}^{10}
  C_{i}({\mu}) O_{i} ({\mu}) \Big]
 +{\rm h.c.},
  \label{eq:eff}
  \end{equation}
  where $p$ $=$ $d$ or $s$,
  $V_{qb} V_{q^{\prime}p}^{\ast}$
  is the product of the Cabibbo-Kobayashi-Maskawa (CKM) matrix
  elements;
  $C_{i}$ are Wilson coefficients, which describe the
  short-distance contributions and are calculated
  perturbatively;
  The explicit expressions of local four-quark operators
  $O_{i}$ are
  \begin{eqnarray}
  O_{1} &=& (\bar{q}_{\alpha}b_{\alpha})_{V-A}
            (\bar{p}_{\beta}q_{\beta}^{\prime})_{V-A},
  \quad \quad \ \ \ \qquad
  O_{2} = (\bar{q}_{\alpha}b_{\beta})_{V-A}
          (\bar{p}_{\alpha}q_{\beta}^{\prime})_{V-A},
  \label{eq:q12} \\
  O_{3} &=& (\bar{p}_{\alpha}b_{\alpha})_{V-A}
             \sum\limits_{p^{\prime}}
            (\bar{p}^{\prime}_{\beta}p^{\prime}_{\beta})_{V-A},
  \quad \ \qquad
  O_{4} = (\bar{p}_{\alpha}b_{\beta})_{V-A}
             \sum\limits_{p^{\prime}}
            (\bar{p}^{\prime}_{\beta}p^{\prime}_{\alpha})_{V-A},
  \label{eq:q34} \\
  O_{5} &=& (\bar{p}_{\alpha}b_{\alpha})_{V-A}
             \sum\limits_{p^{\prime}}
            (\bar{p}^{\prime}_{\beta}p^{\prime}_{\beta})_{V+A},
  \quad \ \qquad
  O_{6} = (\bar{p}_{\alpha}b_{\beta})_{V-A}
             \sum\limits_{p^{\prime}}
            (\bar{p}^{\prime}_{\beta}p^{\prime}_{\alpha})_{V+A},
  \label{eq:q56} \\
  O_{7} &=& (\bar{p}_{\alpha}b_{\alpha})_{V-A}
             \sum\limits_{p^{\prime}}\frac{3}{2}Q_{p^{\prime}}
            (\bar{p}^{\prime}_{\beta}p^{\prime}_{\beta})_{V+A},
  \quad
  O_{8} = (\bar{p}_{\alpha}b_{\beta})_{V-A}
             \sum\limits_{p^{\prime}}\frac{3}{2}Q_{p^{\prime}}
            (\bar{p}^{\prime}_{\beta}p^{\prime}_{\alpha})_{V+A},
  \label{eq:q78} \\
  O_{9} &=& (\bar{p}_{\alpha}b_{\alpha})_{V-A}
             \sum\limits_{p^{\prime}}\frac{3}{2}Q_{p^{\prime}}
            (\bar{p}^{\prime}_{\beta}p^{\prime}_{\beta})_{V-A},
  \quad
  O_{10} = (\bar{p}_{\alpha}b_{\beta})_{V-A}
             \sum\limits_{p^{\prime}}\frac{3}{2}Q_{p^{\prime}}
            (\bar{p}^{\prime}_{\beta}p^{\prime}_{\alpha})_{V-A},
  \label{eq:q90}
  \end{eqnarray}
  where $(\bar{q}_{1}q_{2})_{V{\pm}A}$ $=$
  $\bar{q}_{1}{\gamma}_{\mu}(1{\pm}{\gamma}_{5})q_{2}$,
  ${\alpha}$ and ${\beta}$ are color indices,
  $Q_{p^{\prime}}$ is the electric charge of the quark
  $p^{\prime}$ in the unit of ${\vert}e{\vert}$,
  and $p^{\prime}$ denotes the active quark at the scale
   ${\mu}$ ${\sim}$ ${\cal O}(m_{b})$,
  {\it i.e.}, $p^{\prime}$ $=$ $u$, $d$, $c$, $s$, $b$.

  To obtain the decay amplitudes, the remaining and also
  the most intricate work is how to calculate hadronic
  matrix elements
  ${\langle}PV{\vert}O_{i}{\vert}B^{\ast}{\rangle}$.
  With the factorization approach \cite{Fakirov:1977ta,Bauer:1984zv,Wirbel:1985ji,Bauer:1986bm}
  based on the color transparency mechanism \cite{Bjorken:1988kk,Jain:1995dd},
  in principle, the hadronic matrix element could be
  factorized as
   \begin{equation}
    {\langle}PV{\vert}O_{i}{\vert}B^{\ast}{\rangle}
  =a\,{\langle}P{\vert}J_{\mu}{\vert}B^{\ast}{\rangle}
    {\langle}V{\vert}J^{\mu}{\vert}0{\rangle}
  +b\,{\langle}V{\vert}J_{\mu}{\vert}B^{\ast}{\rangle}
    {\langle}P{\vert}J^{\mu}{\vert}0{\rangle}
  +c\,{\langle}PV{\vert}J_{\mu}{\vert}0{\rangle}
    {\langle}0{\vert}J^{\mu}{\vert}B^{\ast}{\rangle}
   \label{eq:hadonic}.
   \end{equation}

  Due to the unnecessary complexity of hadronic
  matrix element
  ${\langle}V{\vert}J_{\mu}{\vert}B^{\ast}{\rangle}$
  and power suppression of annihilation contributions,
  we only consider one simple scenario where pseudoscalar
  meson pick up the spectator quark in $B^{\ast}$ meson,
  {\it i.e.}, $a$ = 1, $b$ = $0$ and $c$ = 0 in Eq.(\ref{eq:hadonic})
  for the moment.
  Two current matrix elements can be further
  parameterized by decay constants
  and transition form factors,
   \begin{equation}
  {\langle}V(p,{\epsilon}){\vert}
   \bar{q}_{1}{\gamma}_{\mu}q_{2}
  {\vert}0{\rangle} =
  f_{V}m_V{\epsilon}^{\ast}_{\mu},
   \label{eq:fv},
   \end{equation}
   \begin{equation}
  {\langle}P(p_{P}){\vert} \bar{q}{\gamma}_{\mu}b
  {\vert}\bar{B}^{\ast}(p_{B^{\ast}},{\eta}){\rangle} =
   \frac{2V(q^2)}{m_{B^{\ast}}+m_{P}}
  {\varepsilon}_{{\mu}{\nu}{\rho}{\sigma}}
  {\eta}^{\nu}p_{P}^{\rho}p_{B^{\ast}}^{\sigma}
   \label{eq:ffv},
   \end{equation}
   \begin{eqnarray}
   & &
  {\langle}P(p_{P}){\vert} \bar{q} {\gamma}^{\mu}{\gamma}_{5}b
  {\vert}\bar{B}^{\ast}(p_{B^{\ast}},{\eta}){\rangle} =
  i2m_{B^{\ast}}A_0(q^2)\frac{{\eta}{\cdot}q}{q^{2}}q^{\mu}
  \nonumber \\ & & \quad
  +i(m_{P}+m_{B^{\ast}})A_1(q^2) \Big(
  {\eta}^{\mu}-\frac{{\eta}{\cdot}q}{q^{2}} q^{\mu}\Big)
  \nonumber \\ && \quad
  +iA_2(q^2)\frac{{\eta}{\cdot}q}{m_{P}+m_{B^{\ast}}}
   \Big[(p_{B^{\ast}}+p_{P})^{\mu}
  -\frac{(m_{B^{\ast}}^2-m_{P}^2)}{q^2}q^{\mu}\Big]
   \label{eq:ffa},
   \end{eqnarray}
  where ${\epsilon}$ and ${\eta}$ are the
  polarization vector, $f_{V}$ is the decay
  constant of vector meson,
  $V$ and $A_{0,1,2}$ are transition form factors,
  $q$ $=$ $p_{B^{\ast}}$ $-$ $p_{P}$ and
  the sign convention $\epsilon^{0123}=1$.
  Even though some improved approaches, such as
  the QCD factorization \cite{Beneke:1999br,Beneke:2000ry},
  the perturbative QCD scheme \cite{Keum:2000ph,Keum:2000wi} and
  the soft-collinear effective theory ~\cite{Bauer:2000ew,Bauer:2000yr,Bauer:2001ct,Bauer:2001yt},
  are presented to evaluate higher order QCD
  corrections and reduce the renormalization
  scale dependence, the naive factorization (NF)
  approximation is a useful tool of theoretical
  estimation. 
  Because there is no available experimental
  measurement for now,
  the NF approach is good enough to give a preliminary
  analysis, and so adopted in our evaluation.

  With the above definitions, the hadronic matrix
  elements considered here can be decomposed into
  three scalar invariant amplitudes $S_{1,2,3}$,
   \begin{equation}
  {\langle}PV{\vert}O_{i}{\vert}B^{\ast}{\rangle} =
  {\epsilon}^{{\ast}{\mu}}{\eta}^{\nu}
   \Big\{ S_1 g_{{\mu}{\nu}} + S_2
   \frac{ (p_{B^{\ast}}+p_{P})_{\mu}p_{V{\nu}} }{m_{B^{\ast}}m_{V}}
   +iS_3{\varepsilon}_{{\mu}{\nu}{\rho}{\sigma}}
   \frac{2p_{B^{\ast}}^{\rho}p_{P}^{\sigma}}{m_{B^{\ast}}m_{V}} \Big\}
   \label{eq:spd},
   \end{equation}
  where the amplitudes $S_{1,2,3}$ describe the $s$,
  $d$, $p$ wave contributions, respectively, and
  are explicitly written as
  \begin{eqnarray}
  S_1&=&-if_{V}(m_{B^{\ast}}+m_{P})m_{V}A_{1}
  \label{eq:s1}, \\
  S_2&=&-i2f_{V}m_{B^{\ast}}m_{V}^{2}\frac{A_2}{m_{B^{\ast}}+m_{P}}
  \label{eq:s2}, \\
  S_3&=&+i2f_{V}m_{B^{\ast}}m_{V}^{2}\frac{V}{m_{B^{\ast}}+m_{P}}
  \label{eq:s3}.
  \end{eqnarray}
  Alternatively, one can choose the helicity
  amplitudes $H^{\lambda}$ (${\lambda}$ $=$ $0$, $+$, $-$),
  \begin{eqnarray}
  H_{PV}^{0}&=&-S_1x-S_2(x^2-1)
  \label{eq:a0}, \\
  H_{PV}^{\pm}&=&-S_1\,{\pm}\,S_3\sqrt{x^2-1}
  \label{eq:apm},
  \end{eqnarray}
  with
  \begin{equation}
  x\,{\equiv}\,\frac{p_{B^{\ast}}{\cdot}p_{V}}{m_{B^{\ast}}m_{V}}
   =\frac{m_{B^{\ast}}^2-m_{P}^{2}+m_{V}^{2}}{2m_{B^{\ast}}m_{V}}
  \label{eq:x}.
  \end{equation}

  Now, with the formulae given above and  the effective coefficients ${\alpha}_{i}$ defined as
   \begin{equation}
  {\alpha}_{1}=C_{1}+\frac{C_{2}}{N_{c}}, \quad
  {\alpha}_{2}=C_{2}+\frac{C_{1}}{N_{c}}, \quad
  {\alpha}_{4}=C_{4}+\frac{C_{3}}{N_{c}}, \quad
  {\alpha}_{4,EW}=C_{10}+\frac{C_{9}}{N_{c}},
   \end{equation}
  we present the amplitudes of nonleptonic two-body
  $\bar{B}^{\ast}$ decays as follows:
  \begin{itemize}
  \item For $\bar{B}_{q}^{\ast}$ ${\to}$ $D_{q}\bar{D}^{\ast}$ decays (the spectator $q$ $=$ $u$, $d$ and $s$),
   \begin{eqnarray}
  {\cal A}^{\lambda}(\bar{B}_{q}^{\ast}{\to}D_{q}D^{{\ast}-})
  &=&
   H_{DD^{{\ast}-}}^{\lambda} \Big[
   V_{cb} V_{cd}^{\ast} ({\alpha}_{1}+ {\alpha}_{4} + {\alpha}_{4,EW}
    )
  + V_{ub} V_{ud}^{\ast}
    ( {\alpha}_{4} + {\alpha}_{4,EW}
    ) \Big]
   \label{eq:bq-dqdvm}, \\
   {\cal A}^{\lambda}(\bar{B}_{q}^{\ast}{\to}D_{q}D_{s}^{{\ast}-})
  &=&
   H_{DD_{s}^{{\ast}-}}^{\lambda} \Big[
   V_{cb} V_{cd}^{\ast} ({\alpha}_{1}+ {\alpha}_{4} + {\alpha}_{4,EW}
    \Big)
  + V_{ub} V_{ud}^{\ast}
    \Big( {\alpha}_{4} + {\alpha}_{4,EW}
    ) \Big]
   \label{eq:bq-dqdvsm}. 
   \end{eqnarray}

  \item For $\bar{B}_{q}^{{\ast}0}$ ${\to}$ $D_{q}V$ decays (the spectator $q$ $=$ $d$ and $s$, the $V$ $=$ ${\rho}^{-}$ and $K^{{\ast}-}$),
   \begin{eqnarray}
  {\cal A}^{\lambda}(\bar{B}_{q}^{{\ast}0}{\to}D_{q}{\rho}^{-})
  &=&
   H_{D{\rho}^{-}}^{\lambda}
   V_{cb} V_{cd}^{\ast} {\alpha}_{1}
   \label{eq:bq-dqrho}, \\
   {\cal A}^{\lambda}(\bar{B}_{q}^{{\ast}0}{\to}D_{q}K^{{\ast}-})
  &=&
   H_{DK^{{\ast}-}}^{\lambda}
   V_{cb} V_{cs}^{\ast} {\alpha}_{1}
   \label{eq:bq-dqkv}.
   \end{eqnarray}
  \item For $\bar{B}^{\ast}$ ${\to}$ ${\pi}D^{\ast}$ decays,
   \begin{eqnarray}
  {\cal A}^{\lambda}(B^{{\ast}-}{\to}{\pi}^{-}\bar{D}^{{\ast}0})
  &=&
  H^{\lambda}_{{\pi}^{-}\bar{D}^{{\ast}0}}
  V_{ub} V_{cd}^{\ast}{\alpha}_{2}
   \label{Bu-pimDv0}, \\
   \sqrt{2}{\cal A}^{\lambda}(B^{{\ast}-}{\to}{\pi}^{0}D^{{\ast}-})
  &=&
  H^{\lambda}_{{\pi}^{0}D^{{\ast}-}}
  V_{ub} V_{cd}^{\ast}{\alpha}_{1}
   \label{Bu-pi0Dvp}, \\
   \sqrt{2}{\cal A}^{\lambda}(B^{{\ast}-}{\to}{\pi}^{0}D_{s}^{{\ast}-})
  &=&
  H^{\lambda}_{{\pi}^{0}D_{s}^{{\ast}-}}
  V_{ub} V_{cs}^{\ast}{\alpha}_{1}
   \label{Bu-pi0Dvs}, \\
  -\sqrt{2}{\cal A}^{\lambda}(\bar{B}^{{\ast}0}{\to}{\pi}^{0}D^{{\ast}0})
  &=&
  H^{\lambda}_{{\pi}^{0}D^{{\ast}0}}
  V_{cb} V_{ud}^{\ast}{\alpha}_{2}
   \label{Bd-pi0Dv0}, \\
  -\sqrt{2}{\cal A}^{\lambda}(\bar{B}^{{\ast}0}{\to}{\pi}^{0}\bar{D}^{{\ast}0})
  &=&
  H^{\lambda}_{{\pi}^{0}\bar{D}^{{\ast}0}}
  V_{ub} V_{cd}^{\ast}{\alpha}_{2}
   \label{Bd-pi0Dv0b}, \\
  {\cal A}^{\lambda}(\bar{B}^{{\ast}0}{\to}{\pi}^{+}D^{{\ast}-})
  &=&
  H^{\lambda}_{{\pi}^{+}D^{{\ast}-}}
  V_{ub} V_{cd}^{\ast}{\alpha}_{1}
   \label{Bd-pipDvm}, \\
 {\cal A}^{\lambda}(\bar{B}^{{\ast}0}{\to}{\pi}^{+}D_{s}^{{\ast}-})
  &=&
  H^{\lambda}_{{\pi}^{+}D_{s}^{{\ast}-}}
  V_{ub} V_{cs}^{\ast}{\alpha}_{1}
   \label{Bd-pipDvsm}.
   \end{eqnarray}
  \item For $\bar{B}^{\ast}$ ${\to}$ $KD^{\ast}$ decays,
   \begin{eqnarray}
  {\cal A}^{\lambda}(B^{{\ast}-}{\to}K^{-}\bar{D}^{{\ast}0})
  &=&
  H^{\lambda}_{K^{-}\bar{D}^{{\ast}0}}
  V_{ub}V_{cs}^{\ast}{\alpha}_{2}
   \label{bu-kpdv0}, \\
  {\cal A}^{\lambda}(\bar{B}^{{\ast}0}{\to}\bar{K}^{0}\bar{D}^{{\ast}0})
  &=&
  H^{\lambda}_{\bar{K}^{0}\bar{D}^{{\ast}0}}
  V_{ub} V_{cs}^{\ast}{\alpha}_{2}
   \label{bd-kzdv0b}, \\
  {\cal A}^{\lambda}(\bar{B}^{{\ast}0}{\to}\bar{K}^{0}D^{{\ast}0})
  &=&
  H^{\lambda}_{\bar{K}^{0}D^{{\ast}0}}
  V_{cb} V_{us}^{\ast}{\alpha}_{2}
   \label{bd-kzdv0}, \\
  {\cal A}^{\lambda}(\bar{B}_{s}^{{\ast}0}{\to}K^{+}D^{{\ast}-})
  &=&
  H^{\lambda}_{K^{+}D^{{\ast}-}}
  V_{ub} V_{cd}^{\ast}{\alpha}_{1}
   \label{bs-kpdvm}, \\
  {\cal A}^{\lambda}(\bar{B}_{s}^{{\ast}0}{\to}K^{0}\bar{D}^{{\ast}0})
  &=&
  H^{\lambda}_{K^{0}\bar{D}^{{\ast}0}}
  V_{ub} V_{cd}^{\ast}{\alpha}_{2}
   \label{bs-kzdv0b}, \\
  {\cal A}^{\lambda}(\bar{B}_{s}^{{\ast}0}{\to}K^{0}D^{{\ast}0})
  &=&
  H^{\lambda}_{K^{0}D^{{\ast}0}}
  V_{cb} V_{ud}^{\ast}{\alpha}_{2}
   \label{bs-kzdv0}, \\
  {\cal A}^{\lambda}(\bar{B}_{s}^{{\ast}0}{\to}K^{+}D_{s}^{{\ast}-})
  &=&
  H^{\lambda}_{K^{+}D_{s}^{{\ast}-}}
  V_{ub} V_{cs}^{\ast}{\alpha}_{1}
   \label{bs-kpdsm}.
   \end{eqnarray}
  \end{itemize}

  In the rest frame of $\bar{B}^{\ast}$ meson,
  the branching fraction can be written as
  \begin{equation}
 {\cal B}(\bar{B}^{\ast}{\to}PV)
 =\frac{1}{3} \frac{G_{F}^{2}}{2} \frac{1}{8{\pi}}
  \frac{p_{c}}{m^{2}_{B^{\ast}}\Gamma_{\rm tot}(B^{\ast})}
  \sum\limits_{\lambda}
 {\vert}{\cal A}_{\lambda}(\bar{B}^{\ast}{\to}PV){\vert}^{2}
  \label{BR1},
  \end{equation}
  where the momentum of final states is
  \begin{equation}
   p_{c} = \frac{ \sqrt{[m^{2}_{B^{\ast}}-(m_{P}+m_{V})^2]
                        [m^{2}_{B^{\ast}}-(m_{P}-m_{V})^2]} }
                { 2m_{B^{\ast}} }.
  \end{equation}

  The longitudinal, parallel and perpendicular polarization
  fractions are defined as
  \begin{equation}
   f_{L,{\parallel},{\perp}} =
   \frac{ {\vert}{\cal A}_{0,{\parallel},{\perp}}{\vert}^{2} }
        { {\vert}{\cal A}_{0}{\vert}^{2}
         +{\vert}{\cal A}_{\parallel}{\vert}^{2}
         +{\vert}{\cal A}_{\perp}{\vert}^{2} }
  \label{eq:pf},
  \end{equation}
 where ${\cal A}_{\parallel}$ and ${\cal A}_{\bot}$ are parallel and perpendicular amplitudes gotten through
  \begin{equation}   \label{eq:at}
  {\cal A}_{\parallel,\bot} = \frac{1}{\sqrt{2}}({\cal A}_{-}{\pm}{\cal A}_{+})
 \end{equation}
 for $\bar{B}^{\ast}$ decays.
  \section{Numerical Results and Discussion}
  \label{sec03}
  Firstly, we would like to clarify the input parameters
  used in our numerical evaluations.
  For the CKM matrix elements, we adopt the Wolfenstein
  parameterization \cite{Wolfenstein:1983yz} and choose
  the four parameters $A$, $\lambda$, $\rho$ and $\eta$
  as \cite{Charles:2004jd}
  \begin{equation}
  A=0.810^{+0.018}_{-0.024}, \quad
  \lambda=0.22548^{+0.00068}_{-0.00034}, \quad
  \bar{\rho}=0.1453^{+0.0133}_{-0.0073}, \quad
  \bar{\eta}=0.343^{+0.011}_{-0.012},
  \label{eq:ckminput}
  \end{equation}
  with $\bar{\rho}$ = ${\rho}(1-\frac{\lambda^2}{2})$ and
  $\bar{\eta}$ = ${\eta}(1-\frac{\lambda^2}{2})$.

  The decay constants of light vector mesons are \cite{P.Ball:2007jd}
  \begin{equation}
  f_{\rho} = (216 \pm 3 )\,{\rm MeV},\quad
  f_{K^{\ast}} = (220 \pm 5)\,{\rm MeV}
  \label{rhok}.
  \end{equation}
 For the decay constants of $D^{\ast}_{(s)}$ mesons,
  we will take \cite{Lucha:2014xla}
  \begin{equation}
   f_{D^{\ast}} = (252.2 \pm 22.3 \pm 4)\,{\rm MeV}, \quad
   f_{D^{\ast}_{s}} = (305.5 \pm 26.8 \pm 5)\,{\rm MeV},
  \label{dds}
  \end{equation}
  which agree well with the results of the other QCD sum rules
  \cite{Gelhausen:2013wia,Narison:2014ska} and lattice QCD
  with $N_f=2$ \cite{Becirevic:2012ti}.

  \begin{table}[h]
  \caption{The numerical results of form factors within BSW model.}
  \label{tab:Ff}
  \begin{tabular}{c|ccc} \hline \hline
  Transition  &  $V(0)$  &  $A_{1}(0)$  &  $A_{2}(0)$ \\ \hline
  $B^{\ast} \to D$ & $0.76$ & $0.75$&$0.62$ \\
  $B^{\ast} \to K$ & $0.41$ & $0.42$&$0.35$ \\
  $B^{\ast} \to \pi$ &$0.35$ &$0.38$&$0.30$ \\
  $B^{\ast}_{s} \to D_{s}$&$0.72$ &$0.69$&$0.59$ \\
  $B^{\ast}_{s} \to K$&$0.30$ &$0.29$&$0.26$ \\ \hline \hline
  \end{tabular}
  \end{table}

  Besides the decay constants, the $B^{\ast}$ ${\to}$ $P$ transition form factors
   are also essential inputs to
  estimate branching ratios for nonleptonic $B^{\ast}$
  ${\to}$ $PV$ decay.
  In this paper, the Bauer-Stech-Wirbel (BSW) model \cite{Wirbel:1985ji}
  is employed to evaluate the form factors $A_1(0)$,
  $A_2(0)$ and $V(0)$, which could be written as the
  overlap integrals of wave functions of mesons
  \cite{Wirbel:1985ji},
   \begin{eqnarray}
   V^{B^{\ast}{\to}P}(0)
   &=&
   \frac{m_{b}-m_{q}}{m_{B^{\ast}}-m_{P}}J^{B^{\ast}{\to}P}
   \label{eq:ff-v0}, \\
   A^{B^{\ast}{\to}P}_1(0)
   &=&
   \frac{m_{b}+m_{q}}{m_{B^{\ast}}+m_{P}}J^{B^{\ast}{\to}P}
   \label{eq:ff-a1}, \\
   A^{B^{\ast}{\to}P}_2(0)
   &=&
   \frac{2m_{B^{\ast}}}{m_{B^{\ast}}-m_{P}}{A^{B^{\ast}{\to}P}_{0}}(0)
  -\frac{m_{B^{\ast}}+m_{P}}{m_{B^{\ast}}-m_{P}}A_{1}^{B^{\ast}{\to}P}(0)
   \label{eq:ff-a2}, \\
   A^{B^{\ast}{\to}P}_{0}(0)
   &=&
  {\int}d^{2} p_{\perp} {\int}_{0}^{1} dx\,
  {\varphi}_{P}(\vec{p}_{\perp},x)\,{\sigma}_{z}\,
  {\varphi}^{1,0}_{V}(\vec{p}_{\perp},x)
   \label{eq:ff-a0}, \\
   J^{B^{\ast}{\to}P}
   &=&\sqrt{2}{\int}d^{2} p_{\perp} {\int}_{0}^{1} dx\,
   {\varphi}_{P}(\vec{p}_{\perp},x)\,i{\sigma}_{y}\,
   {\varphi}^{1,-1}_{V}(\vec{p}_{\perp},x)\,,
   \label{eq:ff-j}
   \end{eqnarray}
  where $\vec{p}_{\perp}$ is the transverse quark momentum,
  ${\sigma}_{y,z}$ are the Pauli matrix acting on the spin
  indices of the decaying quark, and $m_{q}$ represents
  the mass of nonspectator quark of pseudoscalar meson.
  With the meson wave function ${\varphi}_M(\vec{p}_{\perp},x)$
  as solution of a relativistic scalar harmonic oscillator
  potential \cite{Wirbel:1985ji},
  and ${\omega}$ = 0.4 GeV which determines
  the average transverse quark momentum through
  ${\langle}p^{2}_{\perp}{\rangle}$
  = ${\omega}^2$, we get the numerical results
  of the transition form factors summarized in
  Table \ref{tab:Ff}.
  In our following evaluation,  these numbers
  and $15\%$ of them are used as default inputs
  and uncertainties, respectively.

  To evaluate the branching fractions,
  the total decay widths (or lifetimes)
  $\Gamma_{\rm tot}(B^{\ast})$ are necessary.
  However, there is no available experimental or theoretical
  information for $\Gamma_{\rm tot}(B^{\ast})$ until now.
  Because of the fact that the QED radiative processes
  $B^{\ast}$ ${\to}$ $B{\gamma}$ dominate the
  decays of $B^{\ast}$ mesons, we will take
  the approximation $\Gamma_{\rm tot}(B^{\ast})$
  ${\simeq}$ ${\Gamma}(B^{\ast}{\to}B{\gamma})$.
  The theoretical predictions on
  ${\Gamma}(B^{\ast}{\to}B{\gamma})$
  have been widely evaluated in various
  theoretical models, such as relativistic
  quark model \cite{Goity:2000dk,Ebert:2002xz},
  QCD sum rules \cite{Zhu:1996qy},
  light cone QCD sum rules \cite{Aliev:1995wi},
  light front quark model \cite{Choi:2007se},
  heavy quark effective theory with vector
  meson dominance hypothesis \cite{Colangelo:1993zq}
  or covariant model \cite{Cheung:2014cka}.
  In this paper, the most recent results~\cite{Cheung:2014cka,Choi:2007se}
\begin{eqnarray}
  {\Gamma}(B^{{\ast}+}{\to}B^{+}{\gamma})
  &=& (468^{+73}_{-75})\, {\rm eV}\,, 
  \label{eq:GamBp} \\
  {\Gamma}(B^{{\ast}0}{\to}B^{0}{\gamma})
  &=& (148 \pm 20)\, {\rm eV}\,, 
  \label{eq:GamB0}\\
  {\Gamma}(B_{s}^{{\ast}0}{\to}B_{s}^{0}{\gamma})
  &=& (68 \pm 17)\, {\rm eV}\,,
    \label{eq:GamBs}
  \end{eqnarray}
  which agree with the other theoretical results,
 are approximately treated as
  $\Gamma_{\rm tot}$ in our numerical estimate.

  \begin{table}[ht]
  \caption{The $CP$-averaged branching fractions of nonleptonic
  $B^{\ast}$ weak decays.}
  \label{tab:Br}
  \begin{ruledtabular}
  \begin{tabular}{lccc}
  Decay modes  &  Class  & CKM factors &  $\cal{B}$ \\ \hline
    $B^{{\ast}-}$ ${\to}$ $D^{0}D^{{\ast}-}$
  & T, P, ${\rm P_{ew}}$
  & ${\lambda}^{3}$
  & $(3.9^{+0.2+1.3+0.7}_{-0.2-1.1-0.5}) \times 10^{-10}$ \\
    $\bar{B}^{{\ast}0}$ ${\to}$ $D^{+}D^{{\ast}-}$
  & T, P, ${\rm P_{ew}}$
  & $\lambda^{3}$
  & $(1.2^{+0.1+0.4+0.2}_{-0.1-0.4-0.1}) \times 10^{-9}$ \\
    $B^{{\ast}-}$ ${\to}$ $D^{0}D^{{\ast}-}_{s}$
  & T, P, ${\rm P_{ew}}$
  & $\lambda^{2}$
  & $(1.1^{+0.1+0.4+0.2}_{-0.1-0.3-0.1}) \times 10^{-8}$ \\
    $\bar{B}^{{\ast}0}$ ${\to}$ $D^{+}D^{{\ast}-}_{s}$
  & T, P, ${\rm P_{ew}}$
  & $\lambda^{2}$
  & $(3.4^{+0.2+1.1+0.5}_{-0.2-1.0-0.4}) \times 10^{-8}$ \\
    $\bar{B}^{{\ast}0}_{s}$ ${\to}$ $D^{+}_{s}D^{{\ast}-}$
  & T, P, ${\rm P_{ew}}$
  & $\lambda^{3}$
  & $(2.3^{+0.1+0.8+0.8}_{-0.1-0.7-0.5}) \times 10^{-9}$ \\
    $\bar{B}^{{\ast}0}_{s}$ ${\to}$ $D^{+}_{s}D^{{\ast}-}_{s}$
  & T, P, ${\rm P_{ew}}$
  & $\lambda^{2}$
  & $(6.4^{+0.3+2.1+2.1}_{-0.4-1.9-1.3}) \times 10^{-8}$ \\
    $\bar{B}^{{\ast}0}$ ${\to}$ $D^{+}K^{{\ast}-}$
  & T
  & ${\lambda}^{3}$
  & $(7.6^{+0.4+1.9+1.2}_{-0.4-1.7-0.9}) \times 10^{-10}$ \\
    $\bar{B}^{{\ast}0}_{s}$ ${\to}$ $D^{+}_{s}K^{{\ast}-}$
  & T
  & ${\lambda}^{3}$
  & $(1.5^{+0.1+0.4+0.5}_{-0.1-0.3-0.3}) \times 10^{-9}$ \\
    $\bar{B}^{{\ast}0}$ ${\to}$ $D^{+}{\rho}^{-}$
  & T
  & ${\lambda}^{2}$
  & $(1.3^{+0.1+0.3+0.2}_{-0.1-0.3-0.2}) \times 10^{-8}$ \\
    $\bar{B}^{{\ast}0}_{s}$ ${\to}$ $D^{+}_{s}{\rho}^{-}$
  & T
  & ${\lambda}^{2}$
  & $(2.6^{+0.1+0.6+0.9}_{-0.1-0.6-0.5}) \times 10^{-8}$ \\
    \hline
    $B^{{\ast}-}$ ${\to}$ ${\pi}^{-}\bar{D}^{{\ast}0}$
  & C
  & ${\lambda}^{4}$
  & $(3.1^{+0.2+0.8+0.6}_{-0.2-0.6-0.4}) \times 10^{-14}$ \\
    $B^{{\ast}-}$ ${\to}$ ${\pi}^{0}D^{{\ast}-}$
  & T
  & ${\lambda}^{4}$
  & $(4.6^{+0.4+1.4+0.9}_{-0.4-1.2-0.6}) \times 10^{-13}$ \\
    $\bar{B}^{{\ast}0}$ ${\to}$ ${\pi}^{+}D^{{\ast}-}$
  & T
  & ${\lambda}^{4}$
  & $(2.9^{+0.2+0.9+0.5}_{-0.2-0.8-0.3}) \times 10^{-12}$ \\
    $\bar{B}^{{\ast}0}$ ${\to}$ ${\pi}^{0}D^{{\ast}0}$
  & C
  & ${\lambda}^{2}$
  & $(1.2^{+0.1+0.4+0.2}_{-0.1-0.3-0.1}) \times 10^{-10}$ \\
    $\bar{B}^{{\ast}0}$ ${\to}$ ${\pi}^{0}\bar{D}^{{\ast}0}$
  & C
  & ${\lambda}^{4}$
  & $(4.9^{+0.3+1.4+0.8}_{-0.3-1.2-0.6}) \times 10^{-14}$ \\
    $B^{{\ast}-}$ ${\to}$ ${\pi}^{0}D^{{\ast}-}_s$
  & T
  & ${\lambda}^{3}$
  & $(1.3^{+0.1+0.4+0.2}_{-0.1-0.3-0.2}) \times 10^{-11}$ \\
    $\bar{B}^{{\ast}0}$ ${\to}$ ${\pi}^{+}D^{{\ast}-}_{s}$
  & T
  & ${\lambda}^{3}$
  & $(8.1^{+0.6+2.5+1.3}_{-0.7-2.2-1.0}) \times 10^{-11}$ \\  \hline
    $B^{{\ast}-}$ ${\to}$ $K^{-}\bar{D}^{{\ast}0}$
  & C
  & ${\lambda}^{3}$
  & $(7.4^{+0.6+2.1+1.4}_{-0.6-1.9-1.0}) \times 10^{-13}$ \\
    $\bar{B}^{{\ast}0}$ ${\to}$ $\bar{K}^{0}D^{{\ast}0}$
  & C
  & ${\lambda}^{3}$
  & $(1.7^{+0.1+0.5+0.3}_{-0.1-0.4-0.2}) \times 10^{-11}$ \\
    $\bar{B}^{{\ast}0}$ ${\to}$ $\bar{K}^{0}\bar{D}^{{\ast}0}$
  & C
  & ${\lambda}^{3}$
  & $(2.3^{+0.2+0.7+0.4}_{-0.2-0.6-0.3}) \times 10^{-12}$ \\
    $\bar{B}^{{\ast}0}_{s}$ ${\to}$ $K^{+}D^{{\ast}-}$
  & T
  & ${\lambda}^{4}$
  & $(4.3^{+0.3+1.2+1.4}_{-0.4-1.1-0.9}) \times 10^{-12}$ \\
    $\bar{B}^{{\ast}0}_{s}$ ${\to}$ $K^{0}D^{{\ast}0}$
  & C
  & ${\lambda}^{2}$
  & $(3.6^{+0.2+1.0+1.2}_{-0.2-0.9-0.7}) \times 10^{-10}$ \\
    $\bar{B}^{{\ast}0}_{s}$ ${\to}$ $K^{0}\bar{D}^{{\ast}0}$
  & C
  & ${\lambda}^{4}$
  & $(1.4^{+0.1+0.4+0.5}_{-0.1-0.3-0.3}) \times 10^{-13}$ \\
    $\bar{B}^{{\ast}0}_{s}$ ${\to}$ $K^{+}D^{{\ast}-}_{s}$
  & T
  & ${\lambda}^{3}$
  & $(1.2^{+0.1+0.3+0.4}_{-0.1-0.3-0.2}) \times 10^{-10}$
  \end{tabular}
  \end{ruledtabular}
  \end{table}

  \begin{table}[h]
  \caption{The polarization fractions $f_{L}$ and $f_{\parallel}$
  (in the units of percent).}
  \label{tab:Pol}
  \begin{ruledtabular}
  \begin{tabular}{lcc}
   Decay modes & $f_{L}$ & $f_{\parallel}$ \\ \hline
    $B^{{\ast}-}$ ${\to}$ $D^{0}D^{{\ast}-}$
  & $54^{+2}_{-2}$ & $40^{+2}_{-2}$ \\
    $\bar{B}^{{\ast}0}$ ${\to}$ $D^{+}D^{{\ast}-}$
  & $54^{+2}_{-2}$ & $40^{+2}_{-2}$ \\
    $B^{{\ast}-}$ ${\to}$ $D^{0}D^{{\ast}-}_{s}$
  & $52^{+1}_{-2}$ & $43^{+2}_{-2}$ \\
    $\bar{B}^{{\ast}0}$ ${\to}$ $D^{+}D^{{\ast}-}_{s}$
  & $52^{+1}_{-2}$ & $43^{+2}_{-2}$ \\
    $\bar{B}^{{\ast}0}_{s}$ ${\to}$ $D^{+}_{s}D^{{\ast}-}$
  & $54^{+2}_{-2}$ & $40^{+2}_{-2}$ \\
    $\bar{B}^{{\ast}0}_{s}$ ${\to}$ $D^{+}_{s}D^{{\ast}-}_{s}$
  & $52^{+2}_{-2}$ & $42^{+2}_{-2}$ \\
    $\bar{B}^{{\ast}0}$ ${\to}$ $D^{+}K^{{\ast}-}$
  & $85^{+1}_{-1}$ & $13^{+1}_{-1}$ \\
    $\bar{B}^{{\ast}0}_{s}$ ${\to}$ $D^{+}_{s}K^{{\ast}-}$
  & $85^{+1}_{-1}$ & $13^{+1}_{-1}$ \\
    $\bar{B}^{{\ast}0}$ ${\to}$ $D^{+}\rho^{-}$
  & $88^{+1}_{-1}$ & $10^{+1}_{-1}$ \\
    $\bar{B}^{{\ast}0}_{s}$ ${\to}$ $D^{+}_{s}\rho^{-}$
  & $88^{+1}_{-1}$ & $10^{+1}_{-1}$
  \end{tabular}
  \end{ruledtabular}
  \end{table}

  With the aforementioned values of input parameters
  and the theoretical formula, we present theoretical
  predictions for the observables of $\bar{B}^{\ast}$
  ${\to}$ $D\bar{D}^{\ast}$, $D{\rho}$, $DK^{\ast}$,
  ${\pi}D^{\ast}$, $KD^{\ast}$ decays, in which
  only the (color-suppressed) tree induced decay
  modes are evaluated due to that the branching
  fractions of loop induced decays are very small
  and hardly to be measured soon.
  Our numerical results for the branching fractions and the polarization fractions are summarized in Tables
  \ref{tab:Br} and \ref{tab:Pol}. In Table \ref{tab:Br},  the first,
  second and third theoretical errors are caused
  by uncertainties of the CKM parameters,
  hadronic parameters (decay constants and
  form factors) and total decay widths,
  respectively.
  From Tables \ref{tab:Br} and \ref{tab:Pol},
  it could be found that:

  \begin{enumerate}
  \item [(1)] The hierarchy of branching fractions
  is clear.
  (i)
  The branching fractions of $\bar{B}^{\ast}$
  ${\to}$ ${\pi}D^{\ast}$ and $KD^{\ast}$
  decays are much smaller than the ones of
  $\bar{B}^{\ast}$ ${\to}$ $D\bar{D}^{\ast}$,
  $D{\rho}$ and $DK^{\ast}$ decays,
  which is caused by that the form factors
  of $\bar{B}^{\ast}$ ${\to}$ $D$ transition
  are much larger than those of $\bar{B}^{\ast}$
  ${\to}$ ${\pi}$ and $\bar{B}^{\ast}$
  ${\to}$ $K$ transitions.
  (ii)
  For $\bar{B}^{\ast}$ ${\to}$ $D\bar{D}^{\ast}$,
  $D{\rho}$ and $DK^{\ast}$ decays,
  the hierarchy are induced by two factors:
  one is the CKM factor (see the third column of
  Table \ref{tab:Br}), the other is
  ${\Gamma}_{\rm tot}(B^{{\ast}{\pm}})$ $>$
  ${\Gamma}_{\rm tot}(B^{{\ast}0}_{d})$ $>$
  ${\Gamma}_{\rm tot}(B^{{\ast}0}_{s})$
  [see Eqs.(\ref{eq:GamBp},\ref{eq:GamB0},\ref{eq:GamBs})].

  \item [(2)]
  Besides small form factors, the $\bar{B}^{\ast}$
  ${\to}$ ${\pi}D^{\ast}$, $KD^{\ast}$ decays are
  either color suppressed or the CKM factors
  suppressed, hence have very small branching fractions
  (see Table\ref{tab:Br}) to be hardly measured soon.
  Most of the CKM favored and tree-dominated $\bar{B}^{\ast}$ ${\to}$
  $D\bar{D}^{\ast}$, $D{\rho}$, $DK^{\ast}$ decays,
  enhanced by the relatively large
  $\bar{B}^{\ast}$ ${\to}$ $D$ transition form factors,
  have large branching fractions,  $\gtrsim 10^{-9}$, and thus could be measured in the near future.
  In particular, branching ratios for $\bar{B}_{q}^{\ast}$
  ${\to}$ $D_{q}\bar{D}_{s}^{{\ast}-}$, $D_{q}{\rho}$
  decays can reach up to $10^{-8}$, and hence should
  be sought for with priority and firstly observed
  at the high statistics LHC and Belle-II experiments.
  
The numerical results and above analyses are based on the NF, in which the QCD corrections are not included.  Fortunately, for the color-allowed tree amplitude $\alpha_1$, the NF estimate is stable due to the relatively small QCD corrections~\cite{Beneke:2000ry}. For instance, in $B\to\pi\pi$ and $B\to D^* L$ decays, the results $\alpha_{1}(\pi\pi)= (1.020)_{LO}+(0.018+0.018i)_{NLO}$~\cite{Beneke:1999br} and $\alpha_{1}(D^* L)= (1.025)_{LO}+(0.019+0.013i)_{NLO}$~\cite{Beneke:2000ry} indicate clearly that the ${\cal O}(\alpha_s)$ correction is only about $2\%$ and thus trivial numerically. For the color-suppressed decay modes listed in Tables \ref{tab:Br}, even though the NF estimates would suffer significant ${\cal O}(\alpha_s)$ correction~(about $46\%$ in $B\to\pi\pi$ decays for instance~\cite{Beneke:2009ek} ), they still escape the experimental scope due to their small branching factions $<10^{-9}$,  and thus will not be discussed further. In the following analyses, we will pay our attention only to the color allowed tree-dominated $\bar{B}^{\ast}$ ${\to}$
  $D\bar{D}^{\ast}$, $D{\rho}$, $DK^{\ast}$ decays.
 
  \item [(3)]
For the $B^{{\ast}-}{\to}D^{0}D^{{\ast}-}_{(s)}$ and $\bar{B}^{{\ast}0}{\to}D^{+}D^{{\ast}-}_{(s)}$  decays, the $SU(3)$ flavor symmetry implies the relations
   \begin{eqnarray}
  {\cal A}(B^{{\ast}-}{\to}D^{0}D^{{\ast}-}_{s})
 &{\simeq}&
  {\cal A}(\bar{B}^{{\ast}0}{\to}D^{+}D^{{\ast}-}_{s})
   \label{eq:r05}, \\
  {\cal A}(B^{{\ast}-}{\to}D^{0}D^{{\ast}-})
 &{\simeq}&
  {\cal A}(\bar{B}^{{\ast}0}{\to}D^{+}D^{{\ast}-})
   \label{eq:r06}.
   \end{eqnarray}
  Further considering the theoretical prediciton
  $\Gamma(B^{{\ast}+}{\to}B^{+}{\gamma})/\Gamma(B^{{\ast}0}{\to}B^{0}{\gamma})$
  ${\approx}$ 3 [see Eqs.(\ref{eq:GamBp},\ref{eq:GamB0})] and
  assumption ${\Gamma}_{\rm tot}(B^{\ast})$ ${\simeq}$
  ${\Gamma}(B^{\ast}{\to}B{\gamma})$, one may find the ratio
  \begin{eqnarray}
   R_{\rm du}
   &{\equiv}&
   \frac{{\cal B}(\bar{B}^{{\ast}0}{\to}D^{+}D^{{\ast}-}_{s})}
        {{\cal B}(B^{{\ast}-}{\to}D^{0}D^{{\ast}-}_{s})}
   \simeq
   \frac{\Gamma(B^{{\ast}+}{\to}B^{+}{\gamma})}
        {\Gamma(B^{{\ast}0}{\to}B^{0}{\gamma})}
   \stackrel{theo.}
  {\approx} 3
   \label{eq:rds}, \\
   R_{\rm du}^{\prime}
   &{\equiv}&
   \frac{{\cal B}(\bar{B}^{{\ast}0}{\to}D^{+}D^{{\ast}-})}
        {{\cal B}(B^{{\ast}-}{\to}D^{0}D^{{\ast}-})}
   \simeq R_{\rm du}
   \label{eq:rdd},
  \end{eqnarray}
 which are satisfied in our numerical evaluations. Experimentally,  the first relation  Eq.(\ref{eq:rds}) is
  hopeful to be tested soon due to the large branching fractions.
  For the other potentially detectable $\bar{B}^{{\ast}0}_{d,s}$ ${\to}$ $D\bar{D}^{\ast}$,  $D{\rho}$ and $DK^{\ast}$ decay modes, which branching  fractions ${\gtrsim}$ $10^{-9}$, the U-spin symmetry implies relations
   \begin{eqnarray}
  {\cal A}(\bar{B}^{{\ast}0}{\to}D^{+}D^{{\ast}-})
 &{\simeq}&
  {\cal A}(\bar{B}^{{\ast}0}_{s}{\to}D^{+}_{s}D^{{\ast}-})
   \label{eq:r01}, \\
  {\cal A}(\bar{B}^{{\ast}0}{\to}D^{+}D^{{\ast}-}_{s})
 &{\simeq}&
  {\cal A}(\bar{B}^{{\ast}0}_{s}{\to}D^{+}_{s}D^{{\ast}-}_{s})
  \label{eq:r02}, \\
  {\cal A}(\bar{B}^{{\ast}0}{\to}D^{+}K^{{\ast}-})
 &{\simeq}&
  {\cal A}(\bar{B}^{{\ast}0}_{s}{\to}D^{+}_{s}K^{{\ast}-})
   \label{eq:r03}, \\
  {\cal A}(\bar{B}^{{\ast}0}{\to}D^{+}{\rho}^{-})
 &{\simeq}&
  {\cal A}(\bar{B}^{{\ast}0}_{s}{\to}D^{+}_{s}{\rho}^{-})
   \label{eq:r04}.
   \end{eqnarray}
  As similar to $ R_{\rm du}$, one also could get the ratio and relation
   \begin{eqnarray}
   R_{\rm ds}
   &{\equiv}&
   \frac{{\cal B}(\bar{B}^{{\ast}0}{\to}D^{+}D^{{\ast}-}\,,D^{+}D^{{\ast}-}_{s}\,,D^{+}K^{{\ast}-}\,,D^{+}{\rho}^{-})}
        {{\cal B}(\bar{B}^{{\ast}0}_{s}{\to}D^{+}_{s}D^{{\ast}-}\,,D^{+}_{s}D^{{\ast}-}_{s}\,,D^{+}_{s}K^{{\ast}-}\,,D^{+}_{s}{\rho}^{-})}
   \simeq
   \frac{\Gamma(B^{{\ast}0}_s{\to}B^{0}_s{\gamma})}
        {\Gamma(B^{{\ast}0}{\to}B^{0}{\gamma})}
   \stackrel{theo.}
  {\approx} 2\,,
  \end{eqnarray}
 which is also satisfied in our numerical evaluation.
  So, it is obvious that such ratios $R_{\rm du}$ and $R_{\rm ds}$ are useful for probing
  ${\tau}_{B^{{\ast}0}}/{\tau}_{B^{{\ast}{\pm}}}$ and ${\tau}_{B^{{\ast}0}}/{\tau}_{B^{{\ast}0}_s}$, respectively,
  and further testing the theoretical predictions
  of $\Gamma(B^{{\ast}+}{\to}B^{+}{\gamma})/\Gamma(B^{{\ast}0}{\to}B^{0}{\gamma})$ and $\Gamma(B^{{\ast}0}_s{\to}B^{0}_s{\gamma})/\Gamma(B^{{\ast}0}{\to}B^{0}{\gamma})$
  in various models, such as the results in Refs.
  \cite{Choi:2007se,Ebert:2002xz,Goity:2000dk,Zhu:1996qy,Aliev:1995wi,Colangelo:1993zq,Cheung:2014cka}.

  \item [(4)]
  Besides of branching fraction, the  polarization fractions
  $f_{L,{\parallel},{\perp}}$ are also important observables.
  For the potentially detectable decay modes with branching
  fractions ${\gtrsim}$ $10^{-9}$, our numerical results of
  $f_{L,{\parallel}}$ are summarized in Table \ref{tab:Pol}.
  For the helicity amplitudes ${\cal A}_{\lambda}$,
  the formal hierarchy pattern
  \begin{equation}
 {\cal A}_0:{\cal A}_{-}:{\cal A}_{+}
  = 1 : \frac{{\Lambda}_{\rm QCD}}{m_{b}} :
  \left(\frac{{\Lambda}_{\rm QCD}}{m_{b}}\right)^2
  \label{eq:her}
  \end{equation}
  is naively expected.
  Hence, $\bar{B}^{\ast}$ ${\to}$ $PV$ decays are generally
  dominated by the longitudinal polarization state
  and satisfy $f_{L}$ $\sim$ $1$ $-$ $1/m^{2}_{B^{\ast}}$~\cite{Kagan:2004uw}.
   For $\bar{B}^{\ast}$ ${\to}$ $DV$ $(V=K^{\ast}\,,\rho)$
  decays,  in the heavy-quark limit, the helicity amplitudes $H^{\lambda}$ given by Eqs.~(\ref{eq:a0}) and (\ref{eq:apm})  could be simplified as
  \begin{eqnarray}
  H_{PV}^{0}&\simeq&if_{V}\left[\frac{(m_{B^{\ast}}-m_D)(m_{B^{\ast}}+m_D)^2}{2m_{B^{\ast}}}A_1+\frac{(m_{B^{\ast}}+m_D)(m_{B^{\ast}}-m_D)^2}{2m_{B^{\ast}}}A_2\right]\,
  \label{eq:a0DV}, \\
  H_{PV}^{\pm}&\simeq&if_{V}\left[\frac{(m_{B^{\ast}}-m_D)(m_{B^{\ast}}+m_D)^2}{2m_{B^{\ast}}}A_1\mp \frac{(m_{B^{\ast}}+m_D)(m_{B^{\ast}}-m_D)^2}{2m_{B^{\ast}}}V\right]\nonumber\\
  &&\cdot \frac{2m_{B^{\ast}}m_V}{m_{B^{\ast}}^2-m_D^2}\,
  \label{eq:apmDV}.
  \end{eqnarray}
 The transversity amplitudes could be gotten easily through Eq.~(\ref{eq:at}).  Obviously, due to the helicity suppression factor $2m_{B^{\ast}}m_V/(m_{B^{\ast}}^2-m_D^2)$ $\sim$ $2 m_V/m_{B^{\ast}}$ $\sim$
  ${\Lambda}_{\rm QCD}/m_{b}$, the relation of Eq.(\ref{eq:her}) are roughly
  fulfilled.  As a result, the longitudinal polarization fractions of $\bar{B}^{\ast}$ ${\to}$ $DK^{\ast}$ and $D\rho$ decays are very large~( see Table \ref{tab:Pol} for numerical results).  
  
   It should be noted that above analyses and Eqs.~(\ref{eq:a0DV}) and (\ref{eq:apmDV}) are based on the case of $m_V^2\ll m_{B^{\ast}}^2$, and thus possibly no longer satisfied by $\bar{B}^{\ast}$ ${\to}$ $D\bar{D}^{\ast}$ decays because of the un-negligible vector mass $m_{D^{\ast}}$. In fact, for the $\bar{B}^{\ast}$ ${\to}$ $D\bar{D}^{\ast}$ decays, Eqs. (\ref{eq:a0}) and (\ref{eq:apm}) are simplified as
  \begin{eqnarray}
  H_{PV}^{0}&\simeq&if_{D^*}\left[\frac{(m_{B^{\ast}}+m_D)m_{B^{\ast}}}{2}A_1+\frac{m_{B^{\ast}}}{2(m_{B^{\ast}}+m_D)}(m_{B^{\ast}}^2-4m_{D^*}^2)A_2\right]\,
  \label{eq:a0DD}, \\
  H_{PV}^{\pm}&\simeq&if_{D^*}\left[\frac{(m_{B^{\ast}}+m_D)m_{B^{\ast}}}{2}A_1\mp \frac{m_{B^{\ast}}}{2(m_{B^{\ast}}+m_D)}m_{B^{\ast}} \sqrt{m_{B^{\ast}}^2-4m_{D^*}^2}V\right] \cdot \frac{2m_{D^*}}{m_{B^{\ast}}}\,
  \label{eq:apmDD},
  \end{eqnarray}
  in which, due to $(m_{D^*}^{2}-m_{D}^{2})\ll m_{B^{\ast}}^2$, the approximation $x =\frac{m_{B^{\ast}}^2-m_{D}^{2}+m_{D^*}^{2}}{2m_{B^{\ast}}m_{D^*}}\simeq \frac{m_{B^{\ast}}}{2m_{D^*}}$ is used.
  Because the so-called helicity suppression factor $2m_{D^*}/m_{B^{\ast}}\sim 0.8$ is not small, which is different from the case of  $\bar{B}^{\ast}$ ${\to}$ $DV$ decays, it could be easily found that the relation of Eq.(\ref{eq:her}) doesn't follow. Further considering that $H_{PV}^{\pm}$ are dominated by the term of $A_1$  in Eq.~(\ref{eq:apmDD}) due to its large coefficient, the relation $f_{L}(D\bar{D}^{\ast})\sim f_{\parallel}(D\bar{D}^{\ast})\gg f_{\bot}(D\bar{D}^{\ast})$ could be easily gotten.  
Above analyses and findings are confirmed by our numerical
  results in Table \ref{tab:Pol}, which will
  be tested by future experiments.
  
  \item [(5)]
  As known, there are many interesting phenomena in $B$ meson decays, so it is worthy to explore the possible correlation between $B$ and $B^*$ decays. Taking $\bar{B}^{\ast 0}$ ${\to}$ $D^+ \rho^-$ and $\bar{B}^{ 0}$ ${\to}$ $D^{\ast +} \rho^-$ decays as example, we find that the expressions of their helicity amplitudes  (the former one have be given by Eqs.~(\ref{eq:a0DV}) and (\ref{eq:apmDV}) ) are similar with each other except for the replacements $\bar{B}^{\ast }\leftrightarrow \bar{B}$ and $D \leftrightarrow  D^{\ast }$ everywhere in Eqs.~(\ref{eq:a0DV}) and (\ref{eq:apmDV}). As a result, our analyses in item (4) are roughly suitable for $\bar{B}^0\to D^{\ast +} \rho^-$ decay, and the relation 
  \begin{equation}
 f_{L,\parallel}(\bar{B}^{\ast 0}\to D^+ \rho^-)\simeq f_{L,\parallel}(\bar{B}^0\to D^{\ast +} \rho^-)
 \label{eq:pl}
 \end{equation}
   is generally expected. Interestingly, our prediction $f_{L}^{NF}(\bar{B}^{\ast 0}\to D^+ \rho^-)=(88\pm 1)\%$ is consistent with the result $f_{L}^{\rm WSB}(\bar{B}^0\to D^{\ast +} \rho^-)=87\%$~\cite{Kramer:1992gi}, which is in a good agreement with the experimental data $f_{L}^{\rm exp.}(\bar{B}^0\to D^{\ast +} \rho^-)=  (88.5\pm 1.6\pm 1.2)\%$~\cite{Csorna:2003bw}. The relation Eq.~(\ref{eq:pl}) follows. In addition, the similar correlation as Eq.~(\ref{eq:pl}) also exists in the other $B^*$ and corresponding $B$ decays.

  \end{enumerate}

  \section{Summary}
  \label{sec04}
  In this paper, motivated by the experiments of heavy flavor
  physics at the running LHC and forthcoming SuperKEKB/Belle-II,
  the nonleptonic $\bar{B}^{\ast}$ ${\to}$ $D\bar{D}^{\ast}$,
  $D{\rho}$, $DK^{\ast}$, ${\pi}D^{\ast}$, $KD^{\ast}$ weak
  decay modes are evaluated with factorization approach,
  in which the transition form factors are calculated with
  the BSW model and
  the approximation ${\Gamma}_{\rm tot}(B^{\ast})$
  $\simeq$ $\Gamma(B^{\ast}{\to}B{\gamma})$
  is used to evaluate the branching fractions.
  It is found that:
  (i)
  there are some obvious hierarchy among branching
  fractions, in which the $\bar{B}_{q}^{\ast}$ ${\to}$
  $D_{q}\bar{D}_{s}^{{\ast}-}$ and $D_{q}{\rho}^{-}$
  decays have large branching fractions $\sim$
  $10^{-8}$, and hence should be sought for
  with priority at LHC and Belle-II experiments.
  (ii)
  With the implication of $SU(3)$ (or U-spin)
  flavor symmetry, some useful ratios,  $R_{\rm du}$ and $R_{\rm ds}$, are
  suggested to be verified experimentally.
  (iii)
  The $\bar{B}^{{\ast}0}$ ${\to}$ $DK^{\ast}$
  and $D{\rho}$ decays are dominated by
  the longitudinal polarization states,
  numerically $f_{L}$ $\sim$ [80\%,90\%].
  While, the parallel polarization
  fractions of $\bar{B}^{\ast}$ ${\to}$
  $D\bar{D}^{\ast}$ decays are
  comparable with the longitudinal ones,
  numerically,
  $f_{L}:f_{\parallel}$ $\simeq$ $5:4$. In addition, comparing with $B\to VV$ decays, the relation $ f_{L,\parallel}(\bar{B}^{\ast 0}\to D V)\simeq f_{L,\parallel}(\bar{B}^0\to D V) $ is generally expected.
  These results and findings are waiting
  for confirmation from future LHC and
  Belle-II experiments.

  \section*{Acknowledgments}
  The work is supported by the National Natural
  Science Foundation of China (Grant Nos. 11475055,
  11275057, U1232101 and U1332103).
  Q. Chang is also supported by the Foundation
  for the Author of National Excellent Doctoral
  Dissertation of P. R. China (Grant No. 201317),
  the Program for Science and Technology Innovation
  Talents in Universities of Henan Province
  (Grant No. 14HASTIT036) and the Funding
  Scheme for Young Backbone Teachers of
  Universities in Henan Province
  (Grant No. 2013GGJS-058).

 \end{document}